\documentclass{article}

\usepackage{algos2020}

\usepackage[utf8]{inputenc} 
\usepackage[T1]{fontenc}    
\usepackage{hyperref}       
\usepackage{url}            

\usepackage{booktabs}       
\usepackage{makecell}
\usepackage{multirow}

\usepackage{amsfonts}       
\usepackage{amsmath}
\usepackage{amsthm}
\usepackage{nicefrac}       
\usepackage{microtype}      
\usepackage{lipsum}

\usepackage{rotating}

\theoremstyle{remark}
\newtheorem{remark}{Remark}

\theoremstyle{definition}
\newtheorem{example}{Example}
\newtheorem{definition}{Definition}
\newtheorem{dataset}{Data set}

\usepackage{algorithm}
\usepackage{algorithmic}

\usepackage{tikz}
\usetikzlibrary{shapes}
\usetikzlibrary{shapes.multipart}
\usetikzlibrary{shapes.misc}
\usetikzlibrary{fit}
\usetikzlibrary{arrows}
\usetikzlibrary{patterns}
\usetikzlibrary{decorations.markings}
\tikzset{->-/.style={decoration={
		markings,
		mark=at position .65 with {\arrow{triangle 60}}},postaction={decorate}},
every overlay node/.style={
	draw=white,anchor=north west,
}
}

\newcommand{\mtt}[1]{\text{\texttt{#1}}}
\newcommand{\kg}{\mathcal{K}}

\newcommand{\alphabetn}{\Sigma_V}
\newcommand{\alphabete}{\Sigma_A}
\newcommand{\setn}{V}
\newcommand{\sete}{A}
\newcommand{\srce}{s}
\newcommand{\tgte}{t}
\newcommand{\labeln}{\ell_V}
\newcommand{\labele}{\ell_A}
\newcommand{\ckg}{\mathcal{K}^C}
\newcommand{\calphabetn}{\Sigma_V^C}
\newcommand{\calphabete}{\Sigma_A^C}
\newcommand{\csetn}{V^C}
\newcommand{\csete}{A^C}
\newcommand{\csrce}{s^C}
\newcommand{\ctgte}{t^C}
\newcommand{\clabeln}{\ell_V^C}
\newcommand{\clabele}{\ell_A^C}
\newcommand{\ktock}{\lambda}
\newcommand{\trainset}{N}

\newcommand{\ctrainset}{N^C}

\newcommand{\itraine}[1]{n_#1}
\newcommand{\ctraine}{n^C}
\newcommand{\citraine}[1]{n^C_#1}

\newcommand{\dili}{N^{\text{DILI}}}
\newcommand{\pdili}{N^{\text{DILI}}_\oplus}
\newcommand{\ndili}{N^{\text{DILI}}_\ominus}
\newcommand{\scar}{N^{\text{SCAR}}}
\newcommand{\pscar}{N^{\text{SCAR}}_\oplus}
\newcommand{\nscar}{N^{\text{SCAR}}_\ominus}
\newcommand{\neighborhood}{\mathcal{N}}
\newcommand{\freqneighborhood}{\mathcal{N}_l}
\newcommand{\freqtypes}{\mathcal{T}_{\geq l_{\text{min}}}}
\newcommand{\supportset}{\text{\sc SupportSet}}
\newcommand{\instd}{\mathrm{inst}}

\title{Tackling scalability issues in mining path patterns from knowledge graphs: a preliminary study\thanks{Supported by the \textit{PractiKPharma} project, founded by the French National Research Agency (ANR) under Grant ANR15-CE23-0028, and by the \textit{Snowball} Inria Associate Team.}}

\author{
Pierre Monnin \\
Université de Lorraine, CNRS, Inria, LORIA \\
F-54000 Nancy, France \\
\texttt{pierre.monnin@loria.fr} \\
\And
Emmanuel Bresso\\
Université de Lorraine, CNRS, Inria, LORIA \\
F-54000 Nancy, France \\
\texttt{emmanuel.bresso@loria.fr} \\
\And
Miguel Couceiro \\
Université de Lorraine, CNRS, Inria, LORIA \\
F-54000 Nancy, France \\
\texttt{miguel.couceiro@loria.fr} \\
\And
Malika Smaïl-Tabbone\\
Université de Lorraine, CNRS, Inria, LORIA \\
F-54000 Nancy, France \\
\texttt{malika.smail@loria.fr} \\
\And
Amedeo Napoli\\
Université de Lorraine, CNRS, Inria, LORIA \\
F-54000 Nancy, France \\
\texttt{amedeo.napoli@loria.fr} \\
\And
Adrien Coulet\\
Université de Lorraine, CNRS, Inria, LORIA \\
F-54000 Nancy, France \\
\texttt{adrien.coulet@loria.fr} \\
}

\begin{document}
\maketitle

\begin{abstract}
	Features mined from knowledge graphs are widely used within multiple knowledge discovery tasks such as classification or fact-checking.
	Here, we consider a given set of vertices, called \textit{seed vertices}, and focus on mining their associated \textit{neighboring vertices}, \textit{paths}, and, more generally, \textit{path patterns} that involve classes of ontologies linked with knowledge graphs.
	Due to the combinatorial nature and the increasing size of real-world knowledge graphs, the task of mining these patterns immediately entails scalability issues.
	In this paper, we address these issues by proposing a pattern mining approach that relies on a set of constraints (\textit{e.g.}, support or degree thresholds) and the \textit{monotonicity} property.
	As our motivation comes from the mining of real-world knowledge graphs, we illustrate our approach with PGxLOD, a biomedical knowledge graph.
\end{abstract}


\section{Introduction}

Knowledge graphs~\cite{hogan2020} have a central role in knowledge discovery tasks.
For example, Linked Open Data~\cite{bizerHB09} have been used in all steps of the knowledge discovery process~\cite{ristoskiP16jws}.
In particular, features mined from knowledge graphs have been used in multiple applications such as knowledge base completion~\cite{galarragaHMS14}, explanations~\cite{paulheim12}, or fact-checking~\cite{shiW16}.
Here, we focus on knowledge graphs expressed using Semantic Web standards~\cite{berners2001}.
In this context, \textit{vertices} are either individuals that represent entities of a world (\textit{e.g.}, places, drugs, etc.), literals (\textit{e.g.}, integers, dates, etc.), or classes of individuals (\textit{e.g.}, \texttt{Person}, \texttt{Drug}, etc.).
\textit{Arcs} are defined by triples $\langle \mtt{subject}, \mtt{predicate}, \mtt{object} \rangle$ in the Resource Description Format language.
Such a triple states that the \texttt{subject} is linked to the \texttt{object} by a relationship qualified by the \texttt{predicate} (\textit{e.g.}, \texttt{has-side-effect}, \texttt{has-name}, etc.).
Classes and predicates are defined in ontologies, \textit{i.e.}, formal representations of a domain~\cite{gruber1993translation}, and organized into two hierarchies ordered by the subsumption relation.
In Semantic Web standards, individuals, classes, and predicates are identified by a Uniform Resource Identifier (URI).
We view such a knowledge graph as a directed labeled multigraph $\kg = \left(\alphabetn, \alphabete, \setn, \sete, \srce, \tgte, \labeln, \labele \right)$, where
\begin{itemize}
	\item $\setn$ is the set of vertices.
	\item $\sete$ is the set of arcs connecting vertices through predicates\footnote{Here, we discard literals from $\setn$ and arcs that are incident to literals from $\sete$.}.
	\item $\alphabetn$ is the set of vertex labels, here, their URI\footnote{Hence, $|\alphabetn| = |\setn|$.}.
	\item $\alphabete$ is the set of arc labels, here, URIs of predicates of $\kg$.
	\item $\srce: \sete \to \setn$ (respectively $\tgte: \sete \to \setn$) associates an arc to its source (respectively target) vertex.
	\item $\labeln: \setn \to \alphabetn$ (respectively $\labele: \sete \to \alphabete$) maps a vertex (respectively an arc) to its label.
\end{itemize}
Hence, a triple $\langle \mtt{s}, \mtt{p}, \mtt{o} \rangle$ is represented by two vertices $v_\mtt{s}, v_\mtt{o} \in V$ and an arc $a_{\langle \mtt{s},\mtt{p},\mtt{o}\rangle} \in A$.
The source and target vertices of $a_{\langle \mtt{s},\mtt{p},\mtt{o}\rangle}$ are respectively $v_\mtt{s}$ and $v_\mtt{o}$, \textit{i.e.}, $\srce(a_{\langle \mtt{s},\mtt{p},\mtt{o}\rangle}) = v_\mtt{s}$ and $\tgte(a_{\langle \mtt{s},\mtt{p},\mtt{o}\rangle}) = v_\mtt{o}$.
The labels of $v_\mtt{s}$, $v_\mtt{o}$, and $a_{\langle \mtt{s},\mtt{p},\mtt{o}\rangle}$ are respectively \texttt{s}, \texttt{o}, and \texttt{p}, \textit{i.e.}, $\labeln(v_\mtt{s}) = \mtt{s}$, $\labeln(v_\mtt{o}) = \mtt{o}$, and $\labele(a_{\langle \mtt{s},\mtt{p},\mtt{o}\rangle}) = \mtt{p}$.

In this work, we consider the task of mining features from $\kg$ that are associated with a set of vertices of interest, which we call \textit{seed vertices}.
The set of seed vertices can be defined in intension (\textit{i.e.}, all vertices that instantiate a specified ontology class) or in extension (\textit{i.e.}, by specifying the list of their URIs).
For example, in the biomedical domain, an expert may be interested in mining features associated with vertices that represent drugs causing a specific side effect.
We propose to mine from $\kg$ the three following kinds of features: \textit{neighboring vertices}, \textit{paths}, and \textit{path patterns}.

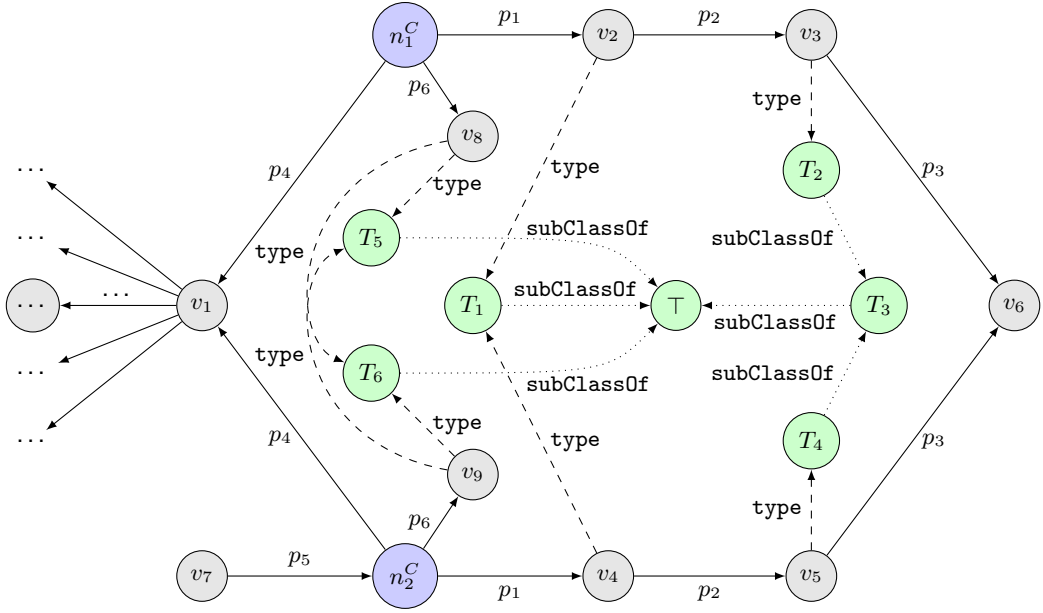
\begin{figure}
	\begin{center}
		\begin{tikzpicture}[align=center,font=\small\sffamily,scale=0.9]
		\node[draw,circle,fill=blue!20] (n1) at (0, 4) {$\citraine{1}$};
		
		\node[draw,circle,fill=blue!20] (n2) at (0, -4) {$\citraine{2}$};
		
		\node[draw,circle,fill=gray!20] (v1) at (-3, 0) {$v_1$};
		\node[draw,circle,fill=gray!20] (v11) at (-5.5, 0) {\dots};
		\node (v12) at (-5.5, 1) {\dots};
		\node (v13) at (-5.5, -1) {\dots};
		\node (v14) at (-5.5, -2) {\dots};
		\node (v15) at (-5.5, 2) {\dots};
		\draw[->, >=latex] (v1) -- node[above] {\dots} (v11);
		\draw[->, >=latex] (v1) -- node[above] {} (v12);
		\draw[->, >=latex] (v1) -- node[above]{} (v13);
		\draw[->, >=latex] (v1) -- node[right] {} (v14);
		\draw[->, >=latex] (v1) -- node[right] {} (v15);
		
		\node[draw,circle,fill=gray!20] (v2) at (3, 4) {$v_2$};
		\node[draw,circle,fill=gray!20] (v3) at (6, 4) {$v_3$};
		
		\node[draw,circle,fill=gray!20] (v4) at (3, -4) {$v_4$};
		\node[draw,circle,fill=gray!20] (v5) at (6, -4) {$v_5$};
		
		\node[draw,circle,fill=gray!20] (v6) at (9, 0) {$v_6$};
		
		\node[draw,circle,fill=gray!20] (v7) at (-3, -4) {$v_7$};
		
		\node[draw,circle,fill=gray!20] (v8) at (1, 2.5) {$v_8$};
		\node[draw,circle,fill=gray!20] (v9) at (1, -2.5) {$v_9$};
		
		\node[draw,circle,fill=green!20] (t1) at (1, 0) {$T_1$};
		\node[draw,circle,fill=green!20] (t2) at (6, 2) {$T_2$};
		
		\node[draw,circle,fill=green!20] (t3) at (7, -0) {$T_3$};
		
		\node[draw,circle,fill=green!20] (t4) at (6, -2) {$T_4$};
		
		\node[draw,circle,fill=green!20] (t5) at (-0.5, 1) {$T_5$};
		\node[draw,circle,fill=green!20] (t6) at (-0.5, -1) {$T_6$};
		
		\node[draw,circle,fill=green!20] (top) at (4, 0) {$\top$};
		
		\draw[->, >=latex] (n1) -- node[left] {$p_4$} (v1);
		\draw[->, >=latex] (n2) -- node[left] {$p_4$} (v1);
		
		\draw[->, >=latex] (n1) -- node[left] {$p_6$} (v8);
		\draw[->, >=latex] (n2) -- node[left] {$p_6$} (v9);
		
		\draw[->, >=latex] (v7) -- node[above] {$p_5$} (n2);
		
		\draw[->, >=latex] (n1) -- node[above] {$p_1$} (v2);
		\draw[->, >=latex] (v2) -- node[above] {$p_2$} (v3);
		\draw[->, >=latex] (v3) -- node[right] {$p_3$} (v6);
		
		\draw[->, >=latex] (n2) -- node[below] {$p_1$} (v4);
		\draw[->, >=latex] (v4) -- node[below] {$p_2$} (v5);
		\draw[->, >=latex] (v5) -- node[right] {$p_3$} (v6);
		
		\draw[->, >=latex,dashed] (v2) -- node[right] {\texttt{type}} (t1);
		\draw[->, >=latex,dashed] (v3) -- node[left] {\texttt{type}} (t2);
		
		\draw[->, >=latex,dashed] (v8) -- node[right] {\texttt{type}} (t5);
		\draw[->, >=latex,dashed] (v8) .. controls (-1.75,2) and (-1.75,-0.5) .. node[left] {\texttt{type}} (t6);
		\draw[->, >=latex,dashed] (v9) -- node[right] {\texttt{type}} (t6);
		\draw[->, >=latex,dashed] (v9) .. controls (-1.75,-2) and (-1.75,0.5) .. node[left] {\texttt{type}} (t5);
		
		\draw[->, >=latex,dotted] (t5) .. controls (3,1) ..  node[above] {\texttt{subClassOf}} (top);
		\draw[->, >=latex,dotted] (t6) .. controls (3,-1) ..  node[below] {\texttt{subClassOf}} (top);
		
		\draw[->, >=latex,dashed] (v4) -- node[right] {\texttt{type}} (t1);
		\draw[->, >=latex,dashed] (v5) -- node[left] {\texttt{type}} (t4);
		
		\draw[->, >=latex,dotted] (t1) -- node[above] {\texttt{subClassOf}} (top);
		\draw[->, >=latex,dotted] (t2) -- node[left] {\texttt{subClassOf}} (t3);
		\draw[->, >=latex,dotted] (t3) -- node[below] {\texttt{subClassOf}} (top);
		\draw[->, >=latex,dotted] (t4) -- node[left] {\texttt{subClassOf}} (t3);
		\end{tikzpicture}
	\end{center}
	\caption{Example of a canonical graph $\ckg$. $\citraine{1}$ and $\citraine{2}$ are canonical seed vertices, all $v_i$ are canonical individuals, and all $T_i$ are canonical ontology classes. Prefixes of URIs were omitted for readability purposes.
		The definition of ``canonical'' is given in Subsection~\ref{subsection:canonicalization}.}
	\label{figure:graph-example}
\end{figure}

\textit{Neighboring vertices} are vertices that can be reached in $\kg$ from at least one seed vertex. 
A neighbor is associated with all seed vertices from which it is reachable.
Its \textit{support} counts such seed vertices.
For example, in the knowledge graph depicted in Figure~\ref{figure:graph-example}, the neighbor $v_6$ is reachable from the seed vertices $\citraine{1}$ and $\citraine{2}$, and thus its support is 2.

\textit{Paths} are sequences of pairs $\xrightarrow{p} e$ that represent an arc labeled by the predicate $p$ incident to an individual $e$.
A path is associated with all seed vertices that root it in $\kg$.
The \textit{support} of a path counts such seed vertices.
For example, the support of $\xrightarrow{p_1} v_2 \xrightarrow{p_2} v_3$ is 1 since only $\citraine{1}$ root it, \textit{i.e.}, $\citraine{1} \xrightarrow{p_1} v_2 \xrightarrow{p_2} v_3$ exists in $\kg$.

More generally, paths may share several characteristics.
For instance, intermediate vertices in paths may instantiate the same ontology classes.
We propose to capture these characteristics by considering \textit{path patterns} in addition to paths.
Path patterns are sequences of pairs $\xrightarrow{p} E$, where $p$ is a predicate and $E$ is either an individual or a class.
Such a pair indicates that an arc labeled by $p$ is incident to \textit{(i)} $E$ if $E$ is an individual or \textit{(ii)} an individual that instantiates $E$ if $E$ is a class.
A path pattern is associated with all seed vertices that root a path captured by the path pattern.
Its \textit{support} counts such seed vertices.
In the example graph depicted in Figure~\ref{figure:graph-example}, $v_2$ instantiates $T_1$, and $v_3$ instantiates $T_2$.
Thus, $\xrightarrow{p_1} v_2 \xrightarrow{p_2} v_3$ is captured by $\xrightarrow{p_1} T_1 \xrightarrow{p_2} v_3$, $\xrightarrow{p_1} v_2 \xrightarrow{p_2} T_2$, and $\xrightarrow{p_1} T_1 \xrightarrow{p_2} T_2$.
Since $T_2$ is a subclass of $T_3$, it is also captured by the pattern $\xrightarrow{p_1} T_1 \xrightarrow{p_2} T_3$.
Note that $\xrightarrow{p_1} T_1 \xrightarrow{p_2} T_3$ also captures $\xrightarrow{p_1} v_4 \xrightarrow{p_2} v_5$, which is rooted by $\citraine{2}$.
Consequently, the support of $\xrightarrow{p_1} T_1 \xrightarrow{p_2} T_3$ is 2.
This illustrates the fact that path patterns may capture additional common characteristics of seed vertices, and thus interestingly complete paths.
Mining these patterns constitutes a challenging task due to the combinatorial nature and the size of real-world knowledge graphs, which naturally entail scalability issues.
For example, $\xrightarrow{p_1} v_2 \xrightarrow{p_2} v_3$ can be generalized by up to 11 path patterns.

This mining task and its inherent scalability issues constitute the main concerns of the present work.
To the best of our knowledge, works available in the literature do not address such issues in knowledge graphs with the adopted granular modeling of path patterns.
However, inspired by existing graph mining works~\cite{aggarwal10}, we propose an Apriori-based approach that alleviates these scalability issues by relying on \textit{(i)} a set of constraints (\textit{e.g.}, support or degree thresholds), \textit{(ii)} the hierarchy of ontology classes, \textit{(iii)} an incremental expansion of paths and patterns, and \textit{(iv)} the monotonic character of the support of paths and patterns.
We provide a reusable implementation on GitHub\footnote{\url{https://github.com/pmonnin/kgpm}}.

The remainder of this paper is organized as follows. 
In Section~\ref{section:related-work}, we outline related works that motivated our proposed approach.
In Section~\ref{section:mining-approach}, we present in details our approach to mine paths and path patterns, and discuss how it tackles scalability issues.
We illustrate our framework in Section~\ref{section:experiments} on PGxLOD, a real-world biomedical knowledge graph~\cite{monninLHRTJNC19}.
We comment on our results as well as indicate directions of future work in Sections~\ref{section:discussion} and~\ref{section:conclusion}.

\section{Related work}
\label{section:related-work}

Path patterns have been widely studied in different settings, for example graph rewriting~\cite{bonfante2018application} and query answering~\cite{barceloLR11}.
Here, we recall some works that tackle the problem of mining path patterns from knowledge graphs in different application contexts.

One major work dealing with feature mining from RDF graphs focuses on graph kernels that count common substructures (\textit{i.e.}, walks, subtrees)~\cite{vriesR13,vriesR15}. 
To avoid an explosion in the number of features, the authors remove patterns with a low or high frequency~\cite{vriesR15}.
Graph features can be used in various tasks such as knowledge base completion.
For example, AMIE~\cite{galarragaTHS13} mines Horn clauses, \textit{i.e.}, conjunction of triples, to predict another triple.
Similarly, in Context Path Model~\cite{stadelmaier2019}, the authors model paths as sequences of predicates between the source and target entities, \textit{i.e.}, $s \xrightarrow{p_1} \xrightarrow{p_2} \cdots \xrightarrow{p_{k-1}}\xrightarrow{p_k} t$.
Here, a triple is predicted based on the paths existing between the involved entities.
Shi and Weninger~\cite{shiW16} also model paths as sequences of predicates but they use them from a fact checking perspective.
They check whether a triple $s \xrightarrow{p} t$ is true by predicting it from a set of learned discriminative paths $\text{\textbf{o}}_s \xrightarrow{p_1} \xrightarrow{p_2} \cdots \xrightarrow{p_{k-1}}\xrightarrow{p_k} \text{\textbf{o}}_t$, where $\text{\textbf{o}}_s$ and $\text{\textbf{o}}_t$ are respectively the set of classes instantiated by $s$ and $t$.
Our framework differs from the previous two as we aim at mining features by exploring the graph from the given seed vertices and we model path patterns using the intermediate entities and the ontology classes that they instantiate.

Our motivation also comes from explainable approaches that rely on the descriptive power of features mined from knowledge graphs.
For example, Explain-a-LOD~\cite{paulheim12} enriches statistical data sets with features from DBpedia.
When correlations can be established between statistics and DBpedia features, these features can be used as explanations for the original statistics.
For example, the quality of living in cities has been correlated with whether these cities are European capitals.
Explain-a-LOD leverages the different outputs of FeGeLOD~\cite{paulheimF12}, some of which corresponding to our approach.
For instance, the so called \textit{relations} $\xrightarrow{p} e$ are paths, whereas the so called \textit{qualified relations} $\xrightarrow{p} t$, where $e$ is replaced by a class $t$ instantiated by $e$, are path patterns.
Alternatively, Vandewiele \textit{et al.}~\cite{vandewiele19} propose to learn a decision tree to classify entities based on paths of a knowledge graph.
The authors suggest that the predictions of their model are explainable as they are obtained by a ``white-box'' model (\textit{i.e.}, the decision tree) that combines interpretable features (\textit{i.e.}, paths from a knowledge graph).
Interestingly, this system considers paths with their intermediate predicates and entities, \textit{i.e.}, \texttt{root} $\xrightarrow{p_1} e_1 \cdots \xrightarrow{p_k} e_k$.
They allow a generalization of both predicates and entities by the use of a wildcard (\texttt{*}).
In our context, this would correspond to generalizing entities by the top level ontology class $\top$.
However, unlike their framework, we do not generalize predicates.
In their study, they focus on paths of the form \texttt{root} $\xrightarrow{\mtt{*}} \mtt{*} \cdots \xrightarrow{\mtt{*}} e$, which somewhat corresponds to extracting neighbors and their distance from seed vertices.

Finally, path patterns are somewhat similar to \textit{generalized association rules}~\cite{srikantA95} and the concept of \textit{raising}~\cite{zhou2008}.
Indeed, both works replace entities in rules by ontology classes to increase the support while preserving a high confidence. 
Inspired by works that prune redundant generalized rules~\cite{dominguesR11}, our approach mines paths and path patterns that are non-redundant and comply with some given constraints.

\section{Towards a scalable approach to mine interesting paths and path patterns}
\label{section:mining-approach}

In this paper, we consider a knowledge graph $\kg$ and a set of seed vertices $\trainset = \left\{\itraine{1}, \itraine{2}, \dots, \itraine{p} \right\} \subseteq \setn$.
The task is to mine neighbors, paths, and path patterns from $\kg$ that are associated with these seed vertices.
For example, given a set of drugs that cause or not a side effect, we aim to mine features that can later be used to classify these drugs.

\begin{figure}
	\begin{center}
		\begin{tikzpicture}[align=center,font=\sffamily,scale=0.9]
		\tikzstyle{every node}=[font=\sffamily\small]
		\draw (-5.9,1.75) -- (-4.75,1.6);
		\draw (-4.3,1.9) -- (-4.75,1.6);
		\draw (-4.75,1.6) -- (-5.1,1);
		\draw (-5.1,1) -- (-4.1,1.3);
		\draw (-5.9,1.75) -- (-6.2,1.3);
		\filldraw (-5.9,1.75) circle (0.1cm);
		\filldraw (-4.75,1.6) circle (0.1cm);
		\filldraw (-4.3,1.9) circle (0.1cm);
		\filldraw (-5.1,1) circle (0.1cm);
		\filldraw (-6.2,1.3) circle (0.1cm);
		\filldraw (-4.1,1.3) circle (0.1cm);
		\node [cloud, draw,cloud puffs=12,cloud puff arc=120, aspect=2, inner ysep=1em,minimum height=1.5cm,minimum width=3cm] (kg-cloud) at (-5,1.5) {};
		
		\node[draw,fill=blue!10] (nodes-sets) at (-5,-1) {Set of seed vertices $\trainset$};
		\node (kg) at (-5,0.45) {Knowledge graph $\kg$};
		
		\node[draw,fill={orange!10},minimum width={4.3cm},minimum height={0.9cm}] (canonicalization) at (1,1.5) {1. Canonicalizing $\kg$};
		
		\node[draw,fill={orange!10},minimum width={4.3cm}] (neighbors-types) at (1,-0.0) {2. Mining interesting \\ neighbors and types};
		
		\node[draw,fill={orange!10},minimum width={4.3cm}] (paths) at (1,-1.5) {3. Mining interesting \\ paths and path patterns};
		
		\node[draw,dashed,fill={yellow!10},minimum width={4cm}] (add-pruning) at (7,1.5) {4. Optional and \\ domain-dependent filtering};
		
		\node[draw,fill={blue!10},minimum width={4cm}] (matrix) at (7,-0.0) {Binary matrix \\ $\mathcal{M} \in \mathbb{B}^{|N| \times |\mathcal{F}|}$};
		
		\draw[->, >=latex] (kg-cloud) -- (canonicalization.west);
		\draw[->, >=latex] (nodes-sets.east) -- (canonicalization.west);
		\draw[->, >=latex] (canonicalization.south) -- (neighbors-types.north);
		\draw[->, >=latex] (neighbors-types.south) -- (paths.north);
		\draw[->, >=latex] (paths.east) -- ++(0.75, 0) -- ++(0, 3) -- (add-pruning.west);
		\draw[->, >=latex] (add-pruning.south) -- (matrix.north);
		\end{tikzpicture}
	\end{center}
	\caption{Main steps to mine a set $\mathcal{F}$ of features (\textit{i.e.}, neighbors, paths, and path patterns) associated with a set $N$ of seed vertices from a knowledge graph $\kg$.
	Step 4 is optional and depends on the application domain.}
	\label{figure:steps-mining}
\end{figure}
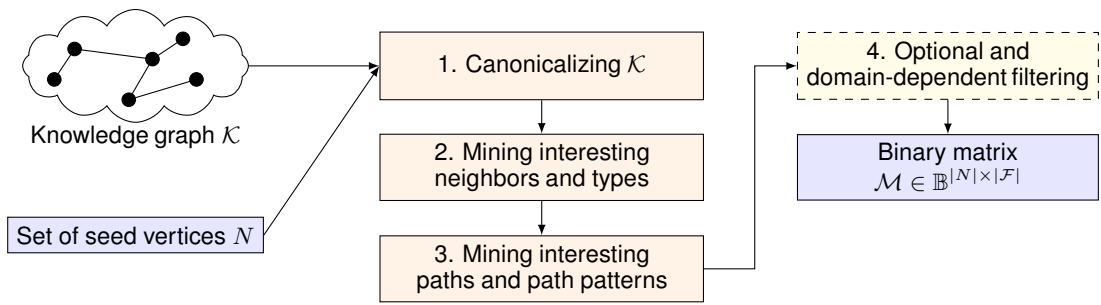

In the following subsections, we propose algorithms to build a binary matrix $\mathcal{M}$ of size $|N| \times |\mathcal{F}|$ from the knowledge graph $\kg$ and the set of seed vertices $N$.
The set $\mathcal{F}$ consists of \textit{interesting} neighbors, paths, and path patterns mined from $\kg$, \textit{i.e.}, neighbors, paths, and path patterns that satisfy the constraints defined in terms of the parameters summarized in Table~\ref{table:mining-parameters}.
These parameters will be detailed in the following subsections.
$\mathcal{M}$ associates a seed vertex $n \in N$ with its features $f \in \mathcal{F}$, \textit{i.e.}, if $\mathcal{M}_{n, f} = \mtt{true}$, then $n$ has feature $f$.
We outline our approach in Figure~\ref{figure:steps-mining} where steps 1, 2, and 3 are mandatory, while step 4 is optional and depends on the application domain.

\begin{table}
	\caption{Parameters that configure the mining of interesting neighbors, paths, and path patterns in a knowledge graph $\kg$.
		Each parameter is associated with a domain and is used in specific steps (see Figure~\ref{figure:steps-mining} for step numbers). 
		Parameter $m$ is specific to the considered application. 
		Here, we illustrate the role of $m$ with the biomedical domain.
	}
	\begin{center}
		\begin{tabular}{lp{2cm}cl}
			\toprule
			Parameter & Domain & Steps & Description \\
			\midrule
			$k$ & $\mathbb{N}^+$ & 2, 3 & Maximum length of paths and path patterns\\
			$t$ & $\mathbb{N}$ & 3 & Maximum level for generalization in class hierarchies\\
			$d$ & $\mathbb{N}$ & 2, 3 & Maximum degree ($u = \mtt{true}$) or out degree ($u = \mtt{false}$) to allow expansion\\
			$l_{\text{min}}$ & $\mathbb{N}$ & 2, 3 & Minimum support for features \\
			$l_{\text{max}}$ & $\mathbb{N}$ & 2, 3 & Maximum support for features \\
			$u$ & $\mathbb{B}$ &  2, 3 & Whether only out arcs ($u = \mtt{false}$) or all arcs ($u = \mtt{true}$) are traversed\\
			$b_{\text{predicates}}$ & List of URIs & 2, 3 & Blacklist of predicates not to traverse \\
			$b_{\text{exp-types}}$ & List of URIs & 2, 3 & Blacklist of classes whose instances are not to reach \\
			$b_{\text{gen-types}}$ & List of URIs & 2, 3 & Blacklist of classes not to use in generalization \\
			$m$ & $\left\{\mtt{none}, \mtt{p}, \mtt{g},\right.$ $\left.\mtt{m}, \mtt{pg}, \mtt{pgm}\right\}$ & 4 & \makecell[tl]{Optional and domain-dependent filtering strategy \\ \textit{Illustrated here with the biomedical domain}} \\
			\bottomrule
		\end{tabular}
	\end{center}
	\label{table:mining-parameters}
\end{table}

\subsection{Canonicalizing $\kg$}
\label{subsection:canonicalization}

The first step of our approach consists in \textit{canonicalizing} the knowledge graph $\kg$, \textit{i.e.}, unifying vertices that represent the same real-world entity.
We use the \textit{canonicalization} word by analogy with the canonicalization of knowledge bases, which consists in unifying equivalent individuals into one~\cite{galarragaHMS14}.
Indeed, in knowledge bases under the Open Information Extraction paradigm, facts and entities can be represented by synonymous terms, which leads to co-existing and equivalent individuals. 
For example, in such knowledge bases, two individuals \textit{Obama} and \textit{Barack Obama} can co-exist.
Similarly, in $\kg$, vertices can be connected through arcs labeled by the \texttt{owl:sameAs} predicate, indicating that these vertices are actually representing the same real-world entity.

Such a situation typically arises when $\kg$ comprises several data sets.
For example, a drug can be represented by two vertices linked by an \texttt{owl:sameAs} arc, resulting from the information extraction of two independent drug-related databases.
Therefore, their merging allows an easy access to the full extent of the knowledge in $\kg$ about the drug they represent.
Such a canonicalization process corresponds to edge contraction in graph theory (\textit{i.e.}, taking graph quotient).
In our framework, it reduces to contracting arcs whose label is the \texttt{owl:sameAs} predicate.

To perform this canonicalization, we must respect the semantics associated with the \texttt{owl:sameAs} predicate, and thus take into account its symmetry and transitivity.
Indeed, an \texttt{owl:sameAs} arc between two vertices either is explicitly stated in $\kg$ or follows from existing arcs and these two properties.
Let us consider a vertex $v$ in $\kg$.
The canonicalization step merges $v$ with all its identical vertices based on \texttt{owl:sameAs} arcs.
To compute this set of vertices, it suffices to compute the connected component of $v$ in the undirected spanning subgraph formed by the \texttt{owl:sameAs} arcs of $\kg$.
Indeed, undirected edges comply with the symmetry of \texttt{owl:sameAs} and connected components comply with the transitivity of \texttt{owl:sameAs}.

As a result, this step takes $\kg$ as input and outputs its \emph{canonical} graph $\ckg = \left(\calphabetn, \calphabete, \csetn, \csete, \csrce, \ctgte, \clabeln, \clabele\right)$.
Similarly to $\kg$, $\ckg$ is a directed labeled multigraph where
\begin{itemize}
	\item $\csetn$ is the set of canonical vertices.
	\item $\csete$ is the set of canonical arcs connecting canonical vertices through predicates.
	\item $\calphabetn$ is the set of canonical vertex labels.
	\item $\calphabete$ is the set of canonical arc labels.
	\item $\csrce: \csete \to \csetn$ (respectively $\ctgte: \csete \to \csetn$) associates a canonical arc to its canonical source (respectively target) vertex.
	\item $\clabeln: \csetn \to \calphabetn$ (respectively $\clabele: \csete \to \calphabete$) maps a canonical vertex (respectively a canonical arc) to its label.
\end{itemize}
Each canonical vertex in $\ckg$ represents a vertex from $\kg$ and all its identical vertices. 
It is possible for a canonical vertex in $\ckg$ to only represent one vertex $v$ from $\kg$ if $v$ has no identical vertices.
This corresponds to creating a surjective mapping $\ktock: \setn \to \csetn$ associating a vertex from $\kg$ to its equivalent canonical vertex in $\ckg$.
Canonical arcs in $\ckg$ are constructed by using $\ktock$ to map the source and target vertices of arcs in $\kg$ to canonical vertices.
Similarly, the set of seed vertices $\trainset$ is mapped to the set of canonical seed vertices, denoted by $\ctrainset$.

\begin{remark}
	\label{remark:relabeling}
	Note that storing URIs has a high memory footprint.
	Thus, in $\ckg$, we use indices in $\mathbb{N}$ instead of URIs to label vertices and arcs, \textit{i.e.}, $\calphabetn \subseteq \mathbb{N}$ and $\calphabete \subseteq \mathbb{N}$. 
	This leads to a reduced memory consumption in subsequent algorithms.
	Each canonical vertex has one unique label, differing from labels of other canonical vertices, \textit{i.e.}, $|\calphabetn| = |\csetn|$.
	This ``relabeling'' is inspired by the work of de Vries and de Rooij~\cite{vriesR13} that use a structure named $pathMap$ to represent a path by an integer.
	We developed our own structure for this relabeling, which we named \texttt{CacheManager}.
\end{remark}

\subsection{Mining interesting neighbors and types}
\label{subsection:mining-neighbors-types}

\subsubsection{Mining interesting neighbors}

Here, we select all vertices that are neighbors of at least one seed vertex in $\ctrainset$ by performing a breadth-first search constrained by parameters $k$, $d$, $u$, $b_{\text{predicates}}$, and $b_{\text{exp-types}}$.
Neighbors are selected by traversing at most $k$ arcs from the seed vertices in $\ctrainset$.
If $u = \mtt{false}$, then only outgoing arcs are traversed; otherwise, all arcs are traversed regardless of their orientation.

However, not all neighboring vertices are of interest.
For example, we want to avoid provenance metadata vertices.
Indeed, they may not constitute discriminative features as they are specific to the vertex they describe.
As we aim to use ontology classes to generate path patterns, we also need to keep the graph exploration over the individuals of $\kg$ and avoid traversing \texttt{rdf:type} arcs.
To this aim, we do not traverse arcs that are labeled by a predicate whose URI or prefix of URI is blacklisted in $b_{\text{predicates}}$.
For example, we blacklist in $b_{\text{predicates}}$ the prefix of the provenance ontology  PROV-O\footnote{\url{http://www.w3.org/ns/prov\#}} and the URI of the \texttt{rdf:type} predicate\footnote{\url{http://www.w3.org/1999/02/22-rdf-syntax-ns\#type}}.

Additionally, we provide a blacklist $b_{\text{exp-types}}$ of URIs or prefixes of classes whose instances must not be reached.
Hence, we do not reach individuals that instantiate directly or indirectly a blacklisted class, by following \texttt{rdf:type} and \texttt{rdfs:subClassOf} arcs.
For example, in a use case of classifying drugs that cause or not a side effect, one may want to avoid neighbors that represent the side effect.
That is why, the ontology class representing the side effect is blacklisted in $b_{\text{exp-types}}$.

When mining neighboring vertices, we may encounter vertices with a high degree, hereafter named \textit{hubs}.
If the graph exploration considered their numerous neighbors, then the size of the selected neighborhood would increase exponentially, thus causing a scalability issue.
Additionally, hub neighbors may not constitute specific and discriminative features.
Indeed, if a hub can be reached from some seed vertices, \textit{i.e.}, appears in their neighborhood, the neighbors of the hub will be reached by the same seed vertices.
That is why, in our approach, we propose to stop the graph exploration at vertices whose degree is strictly greater than parameter $d$\footnote{From a similar assessment, de Vries and de Rooij~\cite{vriesR15} tackle the hub issue by removing edges based on frequency of pairs $(\text{source}, \text{predicate})$ and $(\text{predicate}, \text{target})$.}.

\begin{remark}
	If $u = \mtt{false}$, then the degree of a vertex only counts outgoing arcs, otherwise all arcs are counted.
	The degree does not count arcs whose predicate is blacklisted in $b_{\text{predicates}}$.
	The degree counts arcs incident to a vertex that instantiates a blacklisted class in $b_{\text{exp-types}}$.
\end{remark}

As a result, with $k$, $d$, $u$, $b_{\text{predicates}}$, and $b_{\text{exp-types}}$ fixed, we obtain a set of neighboring vertices, denoted by $\neighborhood(\ctrainset) \subseteq \csetn$. 
Each neighboring vertex $v \in \neighborhood(\ctrainset)$ may only appear in the neighborhood of some seed vertices from $\ctrainset$ w.r.t. the parameters.
Thus, $v$ is associated with these seed vertices, which we indicate by defining the \textit{support set} of $v$.

\begin{definition}[Support set of a neighbor]
	For a given choice of these parameters, the \textit{support set} of a neighboring vertex $v \in \neighborhood(\ctrainset)$ is denoted by $\supportset(v) \subseteq \ctrainset$ and defined as the set of seed vertices from $\ctrainset$ having $v$ as neighbor.
	The support of a neighbor is defined as the cardinal of its support set.
\end{definition}

Note that some vertices in $\neighborhood(\ctrainset)$ are not very discriminative: when they are associated with very few vertices from $\ctrainset$ or nearly all of them.
This motivates the use of parameters $l_{\text{min}}$ and $l_{\text{max}}$ that define the minimum and maximum support for a neighbor to appear in the set  $\mathcal{F}$ of features.
Hence, a neighbor $v \in \neighborhood(\ctrainset)$ constitutes a feature in the output matrix $\mathcal{M}$ if and only if $l_{\text{min}} \leq |\supportset(v)| \leq l_{\text{max}}$.
We denote the set of interesting neighbors to appear in $\mathcal{F}$ by 
\[\freqneighborhood(\ctrainset) = \left\{v \mid v \in \neighborhood(\ctrainset) \text{ and } l_{\text{min}} \leq |\supportset(v)| \leq l_{\text{max}} \right\}.\]
In $\mathcal{M}$, for $\ctraine \in \ctrainset$ and $v \in \freqneighborhood(\ctrainset)$, we have $\mathcal{M}_{\ctraine, v} = \mtt{true}$ if and only if $\ctraine \in \supportset(v)$.

\begin{example}\label{example:freq-neighbors}
	From $\ckg$ in Figure~\ref{figure:graph-example} and with $k = 3$, $d = 4$, $l_{\text{min}} = 2$, $l_{\text{max}} = 3$, $u = \mtt{false}$, $b_{\text{predicates}} = \left\{\mtt{type}, \mtt{subClassOf}\right\}$, and $b_{\text{exp-types}} = \emptyset$, we obtain:
	\begin{align*}
	\neighborhood(\ctrainset) &= \left\{v_1, v_2, v_3, v_4, v_5, v_6, v_8, v_9\right\};\\
	\supportset(v_1) &= \supportset(v_6) = \left\{\citraine{1}, \citraine{2}\right\};\\
	\supportset(v_2) &= \supportset(v_3) = \supportset(v_8) = \left\{\citraine{1}\right\};\\
	\supportset(v_4) &= \supportset(v_5) = \supportset(v_9) = \left\{\citraine{2}\right\};\\
	\freqneighborhood(\ctrainset) &= \left\{v_1, v_6\right\}.
	\end{align*}
	\noindent The graph exploration stops at $v_1$.
	Indeed, it is considered a hub as its degree is greater than $d$. 
	Thus, vertices on the left of Figure~\ref{figure:graph-example} are not explored.
	Because $u = \mtt{false}$, the graph exploration cannot reach $v_7$.
	If $b_{\text{exp-types}} = \left\{T_3\right\}$, then $v_3$ and $v_5$ cannot be traversed, resulting in $\neighborhood(\ctrainset) = \left\{v_1, v_2, v_4, v_8, v_9\right\}$.
\end{example}

\subsubsection{Mining interesting types}

Observe that we can use $\neighborhood(\ctrainset)$ to compute interesting types over the considered neighborhood. 
These interesting types will alleviate a scalability issue arising when building path patterns in Subsection~\ref{subsection:mining-paths-patterns}.
Interesting types must be computed over $\neighborhood(\ctrainset)$ and not $\freqneighborhood(\ctrainset)$ as vertices whose support is below $l_{\text{min}}$ can instantiate interesting types.
As an intuitive example, in Figure~\ref{figure:graph-example}, $T_3$ is associated with both $\citraine{1}$ (because of $v_3$) and $\citraine{2}$ (because of $v_5$).
For $l_{\text{min}} = 2$, $v_3$ and $v_5$ will not be selected as features, however $T_3$ can be used in path patterns.

Parameters $t$ and $b_{\text{gen-types}}$ constrain the ontology classes considered in the construction of path patterns, and thus they are integrated in the computation of interesting types.
Parameter $t$ specifies the maximum level of considered classes in ontology hierarchies.
This level is computed by starting at vertices to generalize and following \texttt{rdf:type} and \texttt{rdfs:subClassOf} arcs.
$t = 0$ only allows to generalize vertices with $\top$, which is considered to be instantiated by all vertices.
For example, $v_3$ can be generalized by $T_2$ and $\top$ if $t = 1$, and by $T_2$, $T_3$, and $\top$ if $t=2$.
Additionally, types used for generalization must not be blacklisted in $b_{\text{gen-types}}$.
This blacklist consists of URIs or prefixes of ontology classes not to be used during the construction of path patterns.
For example, we refrain from considering general classes such as \texttt{pgxo:Drug}\footnote{\url{http://pgxo.loria.fr/Drug}}.

To compute interesting types, we must first compute their support set. 
This motivates the following predicate:
\[\instd(v, T, t, b_{\text{gen-types}}) = \begin{cases}
\mtt{true} \text{ if $v$ instantiates $T$ under parameters $t$ and $b_{\text{gen-types}}$} \\
\mtt{false} \text{ otherwise}
\end{cases}\]
We can then define the support set of an ontology class $T$ as follows:
\begin{definition}[Support set of an ontology class]
	The support set of an ontology class $T$ is the union of support sets of vertices $v \in \neighborhood(\ctrainset)$ that can be generalized by $T$ under parameters $t$ and $b_{\text{gen-types}}$.
	Formally:
	\[\supportset(T) = \bigcup_{\substack{v \in \neighborhood(\ctrainset) \\ \instd(v, T, t, b_{\text{gen-types}})}} \supportset(v) \]
\end{definition}
Finally, the set of interesting types used to build path patterns is defined as 
$\freqtypes = \left\{T \mid l_{\text{min}} \leq |\supportset(T)| \right\}$.

\begin{example} With the same parameters as in Example~\ref{example:freq-neighbors}, we obtain
	\begin{align*}
	\supportset(T_2) &= \left\{\citraine{1}\right\}; & \supportset(T_1) &= \left\{\citraine{1}, \citraine{2}\right\}; & \freqtypes &= \left\{T_1, T_3, T_5, T_6, \top \right\}.
	\end{align*}
\end{example}

\subsection{Mining interesting paths and path patterns}
\label{subsection:mining-paths-patterns}

This step focuses on mining interesting paths and path patterns rooted by seed vertices $\ctraine \in \ctrainset$.
We use the term \textit{path feature} (PF) to indicate a path or a path pattern.

\begin{definition}[Path feature]
	A path feature is a sequence of atomic elements that are pairs $\xrightarrow{p} E$ where $p$ is a predicate and $E$ is either \textit{(i)} an individual (for paths), or \textit{(ii)} an individual or an ontology class (for path patterns).
	The length of a path feature counts the number of its atomic elements.
\end{definition}

\begin{example}
	In Figure~\ref{figure:graph-example}, the path $\xrightarrow{p_1} v_2 \xrightarrow{p_2} v_3 \xrightarrow{p_3} v_6$ can be rooted by $\citraine{1}$ and is of length 3.
	The path pattern $\xrightarrow{p_1} T_1 \xrightarrow{p_2} T_3$ can be rooted by $\citraine{1}$ and $\citraine{2}$ and is of length 2.
\end{example}

Interesting path features are built by a breadth-first expansion starting at vertices in $\ctrainset$.
As previously, $k$ defines the maximum number of arcs traversed.
Hence, path features are of length 1 to $k$.
Observe that a scalability issue arises when mining interesting path features.
Indeed, there may be several paths between two vertices and each vertex in a path can be generalized by several ontology classes.
For instance, for $t = 2$, $\xrightarrow{p_1} v_2 \xrightarrow{p_2} v_3 \xrightarrow{p_3} v_6$ can be generalized by up to 23 path patterns.
We propose a mining procedure that alleviates the scalability issues associated with the mining of path patterns.
This mining procedure relies on the \textit{monotonicity} of the support set of path features, which is defined as follows:

\begin{definition}[Support set of a path feature]
	The support set of a path consists of all vertices from $\ctrainset$ that root it in $\ckg$. Formally,
	\[\supportset(\xrightarrow{p_a} v_a \dots \xrightarrow{p_b} v_b) = \left\{\ctraine \in \ctrainset \mid \ctraine \xrightarrow{p_a} v_a \dots \xrightarrow{p_b} v_b \text{ exists in } \ckg \right\}.\]
	The support set of a path pattern consists of all vertices from $\ctrainset$ that root a path in $\ckg$ that is captured by the path pattern. Formally,
	\begin{multline*}
	\supportset(\xrightarrow{p_a} E_a \dots \xrightarrow{p_b} E_b) = \Big\{\ctraine \in \ctrainset \mid \ctraine \xrightarrow{p_a} v_a \dots \xrightarrow{p_b} v_b \text{ exists in } \ckg \text{ and } \\ \forall v_i, v_i = E_i \text{ or } \instd(v_i, E_i, t, b_{\text{gen-types}})\Big\}.
	\end{multline*}
	The support of a path feature is defined as the cardinal of its support set.
\end{definition}

\begin{example}
	From Figure~\ref{figure:graph-example}, we have $\supportset(\xrightarrow{p_1} v_2 \xrightarrow{p_2} v_3 \xrightarrow{p_3} v_6) = \left\{\citraine{1}\right\}$.
\end{example}

\begin{sidewaysfigure}
	\begin{center}
		\begin{tikzpicture}[align=center,font=\sffamily]
		\tikzstyle{every node}=[font=\small]
		\node (k1) at (0, 11.1) {$h = 1$};
		
		\node[draw,rectangle split,rectangle split parts=2,fill=green!20] (p4v1) at (-2, 4) {$\xrightarrow{p_4} v_1$ \nodepart{two} $\left\{\citraine{1}, \citraine{2} \right\}$};
		\node[draw,rectangle split,rectangle split parts=2,rounded corners=5,pattern color=gray!50,pattern=north east lines] (p4T) at (-2, 2) {$\xrightarrow{p_4}  \top$ \nodepart{two} $\left\{\citraine{1}, \citraine{2} \right\}$};
		\draw[->, >=latex, dashed] (p4v1) -- (p4T);

		\node[draw,rectangle split,rectangle split parts=2, pattern color=gray!50,pattern=north east lines] (p1v2) at (1, 4) {$\xrightarrow{p_1} v_2$ \nodepart{two} $\left\{\citraine{1}\right\}$};
		
		\node[draw,rounded corners=5,rectangle split,rectangle split parts=2] (p1T1) at (0, 2) {$\xrightarrow{p_1} T_1$ \nodepart{two} $\left\{\citraine{1},\citraine{2}\right\}$};
		
		\node[draw,rounded corners=5,rectangle split,rectangle split parts=2,pattern color=gray!50,pattern=north east lines] (p1T) at (2, 2) {$\xrightarrow{p_1} \top$ \nodepart{two} $\left\{\citraine{1},\citraine{2}\right\}$};
		
		\node[draw,rectangle split,rectangle split parts=2, pattern color=gray!50,pattern=north east lines] (p1v4) at (0.9, 0) {$\xrightarrow{p_1} v_4$ \nodepart{two} $\left\{\citraine{2}\right\}$};
		
		\draw[->, >=latex, dashed] (p1v2) -- (p1T1); 
		\draw[->, >=latex, dashed] (p1v2) -- (p1T);
		\draw[->, >=latex, dashed] (p1v4) -- (p1T1);
		\draw[->, >=latex, dashed] (p1v4) -- (p1T);

		\node[draw,rectangle split,rectangle split parts=2, pattern color=gray!50,pattern=north east lines] (p6v8) at (0, 10) {$\xrightarrow{p_6} v_8$ \nodepart{two} $\left\{\citraine{1}\right\}$};
		
		\node[draw,rounded corners=5,rectangle split,rectangle split parts=2,fill=green!20] (p6T5) at (-2, 8) {$\xrightarrow{p_6} T_5$ \nodepart{two} $\left\{\citraine{1},\citraine{2}\right\}$};
		
		\node[draw,rounded corners=5,rectangle split,rectangle split parts=2,fill=green!20] (p6T6) at (0, 8) {$\xrightarrow{p_6} T_6$ \nodepart{two} $\left\{\citraine{1},\citraine{2}\right\}$};
		
		\node[draw,rounded corners=5,rectangle split,rectangle split parts=2,pattern color=gray!50,pattern=north east lines] (p6T) at (2, 8) {$\xrightarrow{p_6} \top$ \nodepart{two} $\left\{\citraine{1},\citraine{2}\right\}$};
		
		\node[draw,rectangle split,rectangle split parts=2, pattern color=gray!50,pattern=north east lines] (p6v9) at (0, 6) {$\xrightarrow{p_6} v_9$ \nodepart{two} $\left\{\citraine{2}\right\}$};
		
		\draw[->, >=latex, dashed] (p6v8) -- (p6T5); 
		\draw[->, >=latex, dashed] (p6v8) -- (p6T6); 
		\draw[->, >=latex, dashed] (p6v8) -- (p6T); 
		\draw[->, >=latex, dashed] (p6v9) -- (p6T5); 
		\draw[->, >=latex, dashed] (p6v9) -- (p6T6);
		\draw[->, >=latex, dashed] (p6v9) -- (p6T);
		
		\draw[dotted] (3.25, -0.8) -- (3.25, 11.3);
		
		\node (k2) at (7.4, 11.1) {$h = 2$};
		
		\node[draw,rectangle split,rectangle split parts=2, pattern color=gray!50,pattern=north east lines] (p1v2p2v3) at (7.4, 10) {$\xrightarrow{p_1} v_2 \xrightarrow{p_2} v_3$ \nodepart{two} $\left\{\citraine{1}\right\}$};
		\draw[->, >=latex] (p1v2) .. controls (2.8,4) and (2.8,10) .. (p1v2p2v3); 
		
		\node[draw,rectangle split,rectangle split parts=2, pattern color=gray!50,pattern=north east lines] (p1v4p2v5) at (7.4, 0) {$\xrightarrow{p_1} v_4 \xrightarrow{p_2} v_5$ \nodepart{two} $\left\{\citraine{2}\right\}$};
		\draw[->, >=latex] (p1v4) -- (p1v4p2v5); 
		
		\node[draw,rounded corners=5,rectangle split,rectangle split parts=2] (p1T1p2T3) at (6.1, 5) {$\xrightarrow{p_1} T_1 \xrightarrow{p_2} T_3$ \nodepart{two} $\left\{\citraine{1}, \citraine{2}\right\}$};
		\draw[->, >=latex, dashed] (p1v2p2v3) .. controls(6.1, 8.5) .. (p1T1p2T3); 
		\draw[->, >=latex, dashed] (p1v4p2v5) .. controls(6.1, 1.5) .. (p1T1p2T3); 
		
		\node[draw,rounded corners=5,rectangle split,rectangle split parts=2,pattern color=gray!50,pattern=north east lines] (p1T1p2T) at (8.7, 5) {$\xrightarrow{p_1} T_1 \xrightarrow{p_2} \top$ \nodepart{two} $\left\{\citraine{1}, \citraine{2}\right\}$};
		\draw[->, >=latex, dashed] (p1v2p2v3) .. controls(8.7, 8.5) .. (p1T1p2T); 
		\draw[->, >=latex, dashed] (p1v4p2v5) .. controls(8.7, 1.5) .. (p1T1p2T); 
		
		\node[draw,rounded corners=5,rectangle split,rectangle split parts=2, pattern color=gray!50,pattern=north east lines] (p1v2p2T3) at (4.8, 7.5) {$\xrightarrow{p_1} v_2 \xrightarrow{p_2} T_3$ \nodepart{two} $\left\{\citraine{1}\right\}$};
		\draw[->, >=latex, dashed] (p1v2p2v3) -- (p1v2p2T3); 
		
		\node[draw,rounded corners=5,rectangle split,rectangle split parts=2, pattern color=gray!50,pattern=north east lines] (p1v2p2T) at (7.4, 7.5) {$\xrightarrow{p_1} v_2 \xrightarrow{p_2} \top$ \nodepart{two} $\left\{\citraine{1}\right\}$};
		\draw[->, >=latex, dashed] (p1v2p2v3) -- (p1v2p2T); 
		
		\node[draw,rounded corners=5,rectangle split,rectangle split parts=2, pattern color=gray!50,pattern=north east lines] (p1T1p2v3) at (10, 7.5) {$\xrightarrow{p_1} T_1 \xrightarrow{p_2} v_3$ \nodepart{two} $\left\{\citraine{1}\right\}$};
		\draw[->, >=latex, dashed] (p1v2p2v3) -- (p1T1p2v3); 
		
		\node[draw,rounded corners=5,rectangle split,rectangle split parts=2, pattern color=gray!50,pattern=north east lines] (p1v4p2T3) at (4.8, 2.5) {$\xrightarrow{p_1} v_4 \xrightarrow{p_2} T_3$ \nodepart{two} $\left\{\citraine{2}\right\}$};
		\draw[->, >=latex, dashed] (p1v4p2v5) -- (p1v4p2T3); 
		
		\node[draw,rounded corners=5,rectangle split,rectangle split parts=2, pattern color=gray!50,pattern=north east lines] (p1v4p2T) at (7.4, 2.5) {$\xrightarrow{p_1} v_4 \xrightarrow{p_2} \top$ \nodepart{two} $\left\{\citraine{2}\right\}$};
		\draw[->, >=latex, dashed] (p1v4p2v5) -- (p1v4p2T); 
		
		\node[draw,rounded corners=5,rectangle split,rectangle split parts=2, pattern color=gray!50,pattern=north east lines] (p1T1p2v5) at (10, 2.5) {$\xrightarrow{p_1} T_1 \xrightarrow{p_2} v_5$ \nodepart{two} $\left\{\citraine{2}\right\}$};
		\draw[->, >=latex, dashed] (p1v4p2v5) -- (p1T1p2v5); 
		
		\draw[dotted] (11.5, -0.8) -- (11.5, 11.3);

		\node (k3) at (15.7, 11.1) {$h = 3$};
		
		\node[draw,rectangle split,rectangle split parts=2, pattern color=gray!50,pattern=north east lines] (p1v2p2v3p3v6) at (15.7, 10) {$\xrightarrow{p_1} v_2 \xrightarrow{p_2} v_3 \xrightarrow{p_3} v_6$ \nodepart{two} $\left\{\citraine{1}\right\}$};
		\draw[->, >=latex] (p1v2p2v3) -- (p1v2p2v3p3v6); 
		
		\node[draw,rectangle split,rectangle split parts=2, pattern color=gray!50,pattern=north east lines] (p1v4p2v5p3v6) at (15.7, 0) {$\xrightarrow{p_1} v_4 \xrightarrow{p_2} v_5 \xrightarrow{p_3} v_6$ \nodepart{two} $\left\{\citraine{2}\right\}$};
		\draw[->, >=latex] (p1v4p2v5) -- (p1v4p2v5p3v6); 
		
		\node[draw,rounded corners=5,rectangle split,rectangle split parts=2, pattern color=gray!50,pattern=north east lines] (p1v2p2v3p3T) at (15.7, 7.5) {$\xrightarrow{p_1} v_2 \xrightarrow{p_2} v_3 \xrightarrow{p_3} \top$ \nodepart{two} $\left\{\citraine{1}\right\}$};
		\draw[->, >=latex, dashed] (p1v2p2v3p3v6) -- (p1v2p2v3p3T);
		
		\node[draw,rounded corners=5,rectangle split,rectangle split parts=2, pattern color=gray!50,pattern=north east lines] (p1v4p2v5p3T) at (15.7, 2.5) {$\xrightarrow{p_1} v_4 \xrightarrow{p_2} v_5 \xrightarrow{p_3} \top$ \nodepart{two} $\left\{\citraine{2}\right\}$};
		\draw[->, >=latex, dashed] (p1v4p2v5p3v6) -- (p1v4p2v5p3T); 
		
		\node[draw,rounded corners=5,rectangle split,rectangle split parts=2,fill=green!20] (p1T1p2T3p3v6) at (13.5, 5) {$\xrightarrow{p_1} T_1 \xrightarrow{p_2} T_3 \xrightarrow{p_3} v_6 $ \nodepart{two} $\left\{\citraine{1}, \citraine{2}\right\}$};
		\draw[->, >=latex, dashed] (p1v2p2v3p3v6) .. controls(13.5, 8) .. (p1T1p2T3p3v6); 
		\draw[->, >=latex, dashed] (p1v4p2v5p3v6) .. controls(13.5, 2) .. (p1T1p2T3p3v6);
		
		\node[draw,rounded corners=5,rectangle split,rectangle split parts=2,pattern color=gray!50,pattern=north east lines] (p1T1p2T3p3T) at (17.9, 5) {$\xrightarrow{p_1} T_1 \xrightarrow{p_2} T_3 \xrightarrow{p_3} \top$ \nodepart{two} $\left\{\citraine{1}, \citraine{2}\right\}$};
		\draw[->, >=latex, dashed] (p1v2p2v3p3v6) .. controls(17.9, 8) .. (p1T1p2T3p3T); 
		\draw[->, >=latex, dashed] (p1v4p2v5p3v6) .. controls(17.9, 2) .. (p1T1p2T3p3T); 
		
		\end{tikzpicture}
	\end{center}
	\caption{Dependency structure used when mining interesting path features from the canonical graph in Figure~\ref{figure:graph-example}.
		Parameters are $k = 3$, $t = 2$, $d = 4$, $u = \mtt{false}$, $b_{\text{predicates}} = \left\{\mtt{type}, \mtt{subClassOf} \right\}$, $b_{\text{exp-types}} = \emptyset$, $b_{\text{gen-types}} = \emptyset$, $l_{\text{min}} = 2$, and $l_{\text{max}}=3$.
		Path features are displayed with their support set.
		Solid arrows represent expansion, dashed arrows represent generalization, rectangles represent paths, and rounded rectangles represent path patterns.
		Hatched rectangles represent discarded path features because of support limits or more specific path features with identical support set.
		Green rectangles represent path features ultimately added to $\mathcal{F}$.
		Blank rectangles represent path features that respect specificity and support constraints but are not in $\mathcal{F}$ because of other features (in green).
		For readability purposes, path features are displayed as the list of their elements instead of indices from $\mathbb{N}$ actually used to save memory.
	}
	\label{figure:path-dependency-structure}
\end{sidewaysfigure}
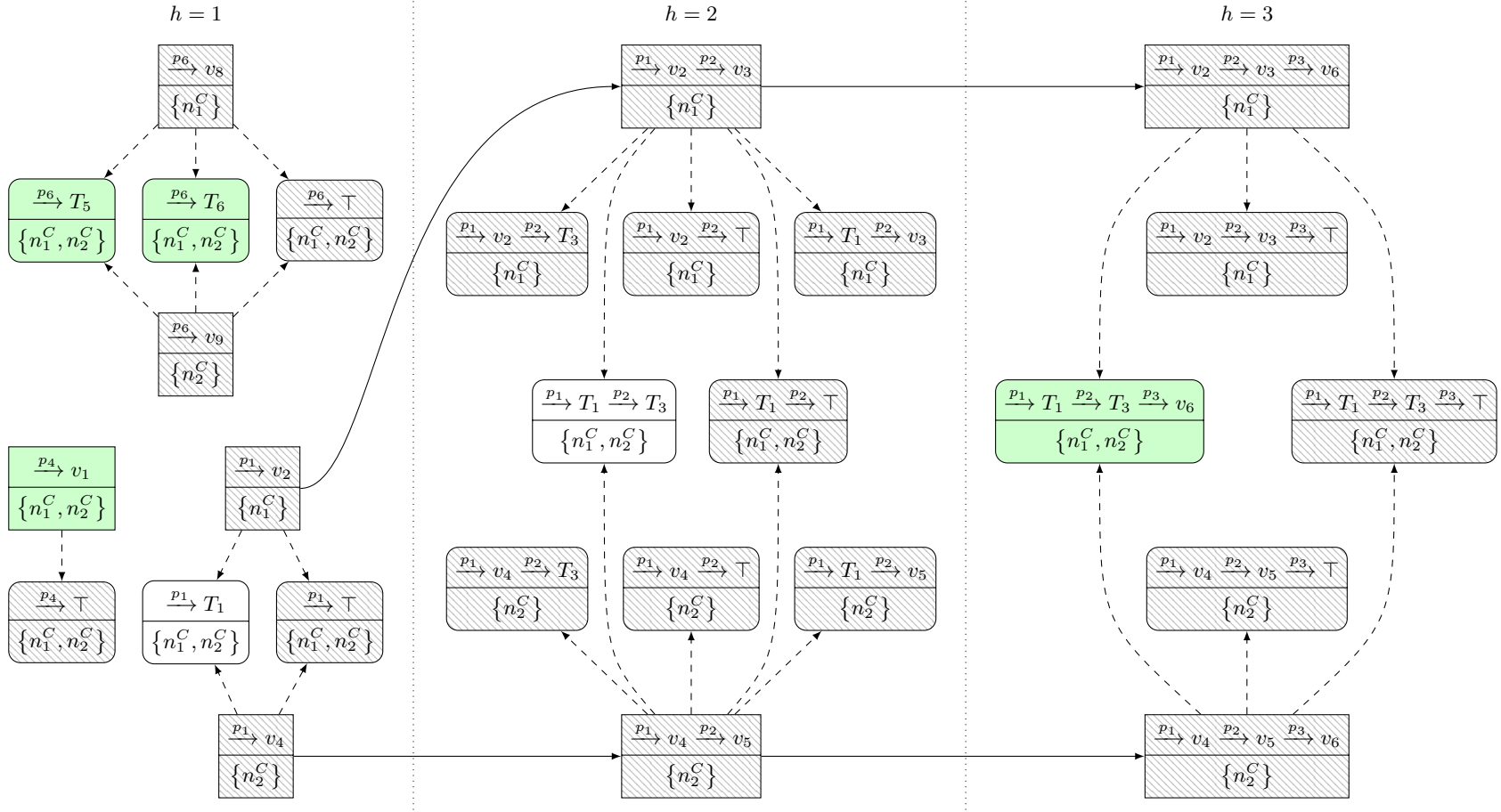

Our approach of mining interesting path features is guided by the \textit{dependency structure} as illustrated in Figure~\ref{figure:path-dependency-structure}.
At first, this structure is empty and it is then augmented at each iteration of Algorithm~\ref{algorithm:mining-path-features}, whose operations are described and illustrated below.

\begin{algorithm}[tb]
	\caption{Mining interesting paths and path patterns}
	\label{algorithm:mining-path-features}
	\textbf{Input}: The canonical knowledge graph $\ckg$, the set of canonical seed vertices $\ctrainset$, the set of interesting types $\freqtypes$\\
	\textbf{Parameters}: $k$, $t$, $d$, $l_\text{min}$, $l_\text{max}$, $u$, $b_\text{predicates}$, $b_\text{exp-types}$, $b_\text{gen-types}$\\
	\textbf{Output}: $\mathcal{F}$ and $\mathcal{M}$ completed with interesting paths and path patterns
	\begin{algorithmic}[1]
		\STATE $h \gets 1$
		\REPEAT
		\STATE Expand paths in $\mathcal{P}_{h}$
		\STATE Generalize expanded paths into path patterns
		\STATE Keep most specific path features
		\STATE Select \textit{(i)} generated path features to add to $\mathcal{F}$ (complete $\mathcal{M}$ accordingly), and \textit{(ii)} paths to add to $\mathcal{P}_{h+1}$
		\STATE $h \gets h + 1$
		\UNTIL{$h > k$ \OR $\mathcal{P}_h = \emptyset$}
	\end{algorithmic}
\end{algorithm}

\begin{remark}
	As introduced in Remark~\ref{remark:relabeling}, there are some hacks to help mitigating some of the scalability drawbacks.
	Storing and manipulating a path feature as a list of elements has a high memory footprint.
	Thus, such a list is only stored once in our \texttt{CacheManager} structure and the returned index (from $\mathbb{N}$) is used in mining algorithms.
\end{remark}

\subsubsection{Expand paths in $\mathcal{P}_h$}

Each path $P \in \mathcal{P}_h$\footnote{$\mathcal{P}_h$ is explained in Subsection~\ref{subsection:feature-selection}.} is expanded with pairs $\xrightarrow{p_e} v_{e}$ that are chosen in the neighborhood of the last individual of $P$.
This choice is constrained by parameters $k$, $d$, $u$, $b_{\text{predicates}}$, and $b_{\text{exp-types}}$ as in Subsection~\ref{subsection:mining-neighbors-types}\footnote{Additionally, to avoid loops, $P$ can only be expanded at iteration $h$ with individuals $v_{e}$ such that there exists at least one seed vertex in $\supportset(P)$ whose shortest distance to $v_{e}$ is $h$.}.
In the first iteration, the neighborhood of seed vertices in $\ctrainset$ is used.
If no path in $\mathcal{P}_h$ can be expanded, \textit{i.e.}, the neighborhood of their last vertex does not contain reachable vertices under the constraints, then Algorithm~\ref{algorithm:mining-path-features} ends.

\begin{example}
	From the graph in Figure~\ref{figure:graph-example}, the first expansion generates the following paths: $\xrightarrow{p_4} v_1$, $\xrightarrow{p_1} v_2$, $\xrightarrow{p_1} v_4$, $\xrightarrow{p_6} v_8$, and $\xrightarrow{p_6} v_9$.
	In the second expansion, $\xrightarrow{p_4} v_1$ is not expanded as $v_1$ is a hub under $d = 4$.
	Since their respective neighborhood does not contain reachable vertices, $\xrightarrow{p_6} v_8$ and $\xrightarrow{p_6} v_9$ are also not expanded.
	The expansion of $\xrightarrow{p_1} v_2$ generates $\xrightarrow{p_1} v_2 \xrightarrow{p_2} v_3$ whereas the expansion of $\xrightarrow{p_1} v_4$ generates $\xrightarrow{p_1} v_4 \xrightarrow{p_2} v_5$.
\end{example}

\subsubsection{Generalize expanded paths into path patterns}

Let $P_e$ be the expansion of $P \in \mathcal{P}_h$ as previously described, \textit{i.e.}, $P_e = P \xrightarrow{p_e} v_e$.
We generalize $P_e$ by:
\begin{itemize}
	\item Generating patterns $P \xrightarrow{p_e} T$ with types $T \in \freqtypes$ for which the predicate $\instd(v_{e}, T, t, b_{\text{gen-types}})$ is verified.
	\item Retrieving from the dependency structure the path patterns that generalize $P$\footnote{Not all generated path patterns remain in the dependency structure at the end of an iteration, see Subsections~\ref{subsection:mspp} and~\ref{subsection:feature-selection}.} and expanding them with $\xrightarrow{p_e} v_e$, and $\xrightarrow{p_e} T$ for all $T \in \freqtypes$ for which the predicate $\instd(v_{e}, T, t, b_{\text{gen-types}})$ is verified.
\end{itemize}
Intuitively, this generalization operation allows to expand path patterns.

\begin{example}
	In the first iteration, $\xrightarrow{p_1} v_2$ is generalized by $\xrightarrow{p_1} T_1$ and $\xrightarrow{p_1} \top$.
	As we will see, $\xrightarrow{p_1} \top$ is not kept in the dependency structure at the end of the first iteration (see Subsections~\ref{subsection:mspp} and~\ref{subsection:feature-selection}).
	In the second iteration, $\xrightarrow{p_1} v_2$ expands into $\xrightarrow{p_1} v_2 \xrightarrow{p_2} v_3$.
	In the dependency structure, we retrieve $\xrightarrow{p_1} T_1$ as the path pattern generalizing $\xrightarrow{p_1} v_2$, which is expanded into $\xrightarrow{p_1} T_1 \xrightarrow{p_2} v_3$, $\xrightarrow{p_1} T_1 \xrightarrow{p_2} T_3$, and $\xrightarrow{p_1} T_1 \xrightarrow{p_2} \top$.
\end{example}

We only generalize paths with types $T \in \freqtypes$ to avoid the generation an important number of uninteresting path patterns that would then be discarded, thus reducing the memory footprint.
Indeed, by definition, if $T \not\in \freqtypes$, then $|\supportset(T)| < l_{\text{min}}$.
Additionally, given a path feature $P$, $|\supportset(P)| \leq \min_{E \in P} |\supportset(E)|$, where $E$ can be a class or an individual involved in $P$.
Thus, if $T \not\in \freqtypes$ is used in a path pattern $P$, we would have $|\supportset(P)| < l_{\text{min}}$, therefore generating an uninteresting path pattern that would be discarded later.

\subsubsection{Keep most specific path features}
\label{subsection:mspp}

Inspired by works that prune redundant generalized rules during their generation~\cite{dominguesR11}, we keep only the ``most specific'' path patterns among those that have the same support set.
\begin{definition}[More specific path pattern]
	\label{definition:mspp}
	A path pattern $P_1$ is \textit{more specific} than another path pattern $P_2$ if every atomic element of $P_1$ is more specific than the atomic element of $P_2$ at the same position.
	An atomic element $\xrightarrow{p_1} E_1$ is more specific than another atomic element $\xrightarrow{p_2} E_2$ if and only if:
	\begin{itemize}
		\item[(i)] $p_1 = p_2$, \textit{i.e.}, both atomic elements involve the same predicate\footnote{It is noteworthy that we do not consider the hierarchy of predicates in this work.}, and
		\item[(ii)] $E_1$ is more specific than $E_2$\footnote{A class is more specific than all its super-classes and an individual is more specific than all classes it instantiates.}.
	\end{itemize}
\end{definition}
When path patterns have the same support set, keeping the most specific ones remove redundant generalizations, thus reducing their number and the computational burden.
Additionally, we ensure a high descriptive power because the most specific paths are the most descriptive. Intuitively, a path pattern involving a class $T$ is less descriptive than another pattern involving a subclass or an instance of the class.
However, keeping the most specific path patterns is computationally expensive, which led us to propose the following computational procedure.

We notice that the support set of a path pattern is the union of the support sets of the paths it generalizes.
Therefore, we discard path patterns that generalize only one path.
Indeed, such path patterns have the same support set as their original path and are more general, by definition.

\begin{example}
	For $h = 3$, we discard the following path patterns: $\xrightarrow{p_1} v_2 \xrightarrow{p_2} v_3 \xrightarrow{p_3} \top$ and $\xrightarrow{p_1} v_4 \xrightarrow{p_2} v_5 \xrightarrow{p_3} \top$.
\end{example}

\begin{figure}
	\begin{center}
		\begin{tikzpicture}[align=center,font=\sffamily]
		\node[minimum width={2cm}, anchor=center] (k1) at (1.6, 7.5) {\strut $h = 1$ \\ Support set $\left\{\citraine{1}, \citraine{2}\right\}$};
		
		\node[draw, minimum height={0.8cm}, minimum width={0.8cm}, align=center] (k1p1) at (0, 5.8) {\strut$p_1$};
		\node[draw, minimum height={0.8cm}, minimum width={0.8cm}, align=center] (v1p1) at (0.8, 5.8){\strut};
		\node[draw, minimum height={0.8cm}, minimum width={0.8cm}, align=center] (k1p6) at (0, 5) {\strut$p_6$};
		\node[draw, minimum height={0.8cm}, minimum width={0.8cm}, align=center] (v1p6) at (0.8, 5){\strut};
		
		\node[draw, minimum height={0.8cm}, minimum width={0.8cm}, align=center] (v2p1) at (3, 6.2) {\strut$\left\{\top\right\}$};
		\node[draw, minimum height={0.8cm}, minimum width={0.8cm}, align=center] (v2p6) at (3, 4.6) {\strut$\left\{T_5, T_6\right\}$};
		
		\draw[->, >=latex] (v1p1.center) -- (v2p1.west);
		\draw[->, >=latex] (v1p6.center) -- (v2p6.west);
	
		\draw[dotted] (4, 4) -- (4, 8);
		
		\node (k2) at (9.3, 7.5) {$h = 2$ \\ Support set $\left\{\citraine{1}, \citraine{2}\right\}$};
		
		\node[draw, minimum height={0.8cm}, minimum width={0.8cm}, align=center] (k3p1) at (5, 5.8) {\strut$p_1$};
		\node[draw, minimum height={0.8cm}, minimum width={0.8cm}, align=center] (v3p1) at (5.8, 5.8){\strut};
		
		\node[draw, minimum height={0.8cm}, minimum width={0.8cm}, align=center] (k3T1) at (7.6, 5.8) {\strut$T_1$};
		\node[draw, minimum height={0.8cm}, minimum width={0.8cm}, align=center] (v3T1) at (8.4, 5.8){\strut};
		
		\draw[->, >=latex] (v3p1.center) -- (k3T1.west);
		
		\node[draw, minimum height={0.8cm}, minimum width={0.8cm}, align=center] (k3p2) at (10.2, 5.8) {\strut$p_2$};
		\node[draw, minimum height={0.8cm}, minimum width={0.8cm}, align=center] (v3p2) at (11, 5.8){\strut};
		
		\draw[->, >=latex] (v3T1.center) -- (k3p2.west);
		
		\node[draw, minimum height={0.8cm}, minimum width={0.8cm}, align=center] (k3T) at (12.8, 5.8) {\strut$\left\{T_3\right\}$};
		
		\draw[->, >=latex] (v3p2.center) -- (k3T.west);
		\end{tikzpicture}
	\end{center}
	\caption{Prefix tree used when computing most specific path patterns during the first and second iterations considering the support set $\left\{\citraine{1}, \citraine{2}\right\}$. 
		This structure is reset at each expansion and for each support set.
		For $h = 1$, when considering $\xrightarrow{p_1} T_1$, the structure will evolve: $\xrightarrow{p_1} \top$ will be removed and replaced by the considered path.
		For $h = 3$, when considered, $\xrightarrow{p_1} T_1 \xrightarrow{p_2} \top$ will be discarded as more general than $\xrightarrow{p_1} T_1 \xrightarrow{p_2} T_3$.
	}
	\label{figure:mspp-prefix-tree}
\end{figure}
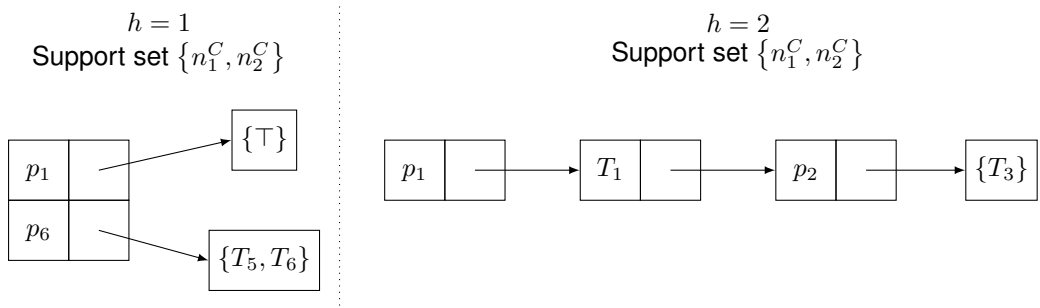

However, there may exist path patterns that generalize several more specific path features while having the same support set. 
Such path patterns should also be discarded.
\begin{example}
	For $h = 1$, $\xrightarrow{p_1} \top$ shares the same support set as the more specific path pattern $\xrightarrow{p_1} T_1$ and thus should be discarded.
\end{example}
To efficiently discard path patterns, we avoid computing their whole hierarchy.
Instead, we focus on retaining only the most specific ones in the \textit{prefix tree} depicted in Figure~\ref{figure:mspp-prefix-tree}.
This prefix tree is incrementally augmented and stores the most specific path patterns for a specific iteration and support set.
In this tree, individuals/classes and predicates involved in path patterns are indexed separately.
Thus, its depth is twice the length of path patterns of the current iteration.

The prefix tree enables an efficient storage and selection of the most specific path patterns.
Indeed, let $P$ be a path pattern to be compared with those already stored.
A breadth-first traversal is performed to detect more specific patterns than $P$: we only traverse identical or more specific elements according to Definition~\ref{definition:mspp}.
At any depth, if such elements cannot be found, then $P$ is one of the most specific patterns and the traversal stops.
On the contrary, if the traversal reaches a leaf containing more specific elements, then there is a pattern more specific than $P$ with the same support set. 
Consequently, $P$ is discarded and removed from the dependency structure.

If $P$ is to be stored, another breadth-first traversal is performed by considering identical or more general elements than the ones in $P$.
When the traversal reaches a leaf, it means that more general patterns than $P$ are currently stored and have the same support set.
These are removed from the prefix tree before storing $P$. 
They are also removed from the dependency structure.

\begin{remark}
	As the prefix tree relies on associative arrays and sets, the computational cost of traversal, insertion, and removal is reduced.
	Additionally, resetting the tree at each expansion and each support set reduces the number of patterns to traverse and thus the global computational cost.
\end{remark}

\subsubsection{Select generated path features to add to $\mathcal{F}$, and paths to add to $\mathcal{P}_{h+1}$}
\label{subsection:feature-selection}

In this subsection, we determine \textit{(i)} which paths and path patterns should be added as features in $\mathcal{F}$, and consequently, \textit{(ii)} which paths to add to $\mathcal{P}_{h+1}$, \textit{i.e.}, to expand during the next iteration.

A path feature $P$ can be added in $\mathcal{F}$ if:
\begin{itemize}
	\item[(C1)] $l_{\text{min}} \leq \supportset(P) \leq l_{\text{max}}$,
	\item[(C2)] Prefixes of $P$ with the same support set do not already exist in $\mathcal{F}$,
	\item[(C3)] If $P$ is a path pattern, it does not generalize a path with the same support set.
\end{itemize}
When $P$ is in $\mathcal{F}$, the output binary matrix $\mathcal{M}$ verifies $\mathcal{M}_{\ctraine, P} = \mtt{true}$ for all $\ctraine \in \supportset(P)$.

\begin{remark}
	Note that (C2) allows to focus on shorter paths in $\mathcal{F}$.
	However, (C2) is not applied if $P$ ends with an individual and there exist a prefix of $P$ in $\mathcal{F}$ that ends with a class.
	$P$ is considered more descriptive than its prefix, because of the individual in the last position.
	Hence, we add $P$ in $\mathcal{F}$ and remove its prefix.
	For example, in Figure~\ref{figure:path-dependency-structure}, $\xrightarrow{p_1} T_1 \xrightarrow{p_2} T_3$ is replaced by $\xrightarrow{p_1} T_1 \xrightarrow{p_2} T_3 \xrightarrow{p_3} v_6$ in $\mathcal{F}$ for $h = 3$.
\end{remark}

\begin{remark}
	(C3) is motivated as the path is more specific and thus more descriptive than $P$ and should be added instead of $P$.
\end{remark}

We also select the paths and path patterns to expand during the next iteration.
To reduce their number, we rely on the $l_\text{min}$ constraint and the \textit{monotonicity} of the support set.
It is clear that, for a path feature $P$, we have $|\supportset(P)| \leq \min_{E \in P} |\supportset(E)|$, where $E$ can be a class or an individual involved in $P$.
Thus, when expanding a path feature, its support set remains identical (for paths and path patterns) or decreases (for path patterns).

Consequently, we add to $\mathcal{P}_{h+1}$ paths whose expansion may generate path features complying with the $l_\text{min}$ constraint, \textit{i.e.}, paths with a support greater than $l_{\text{min}}$ or paths that are generalized by a pattern with a support greater than $l_\text{min}$.
This monotonicity property also lets us remove from the dependency structure patterns whose support is lower than $l_\text{min}$.
Indeed, such patterns cannot be used during the generalization step of the next iteration since they will inevitably generate a pattern whose support is smaller than $l_\text{min}$.
As a result, the monotonicity property does entail a reduction in the number of paths and path patterns considered in the next iteration, thus reducing the computational cost.

\begin{example}
	For example, in the first iteration, the path $\xrightarrow{p_1} v_2$ is added to $\mathcal{P}_{2}$ as it is generalized by $\xrightarrow{p_1} T_1$ whose support is greater than $l_\text{min} = 2$.
	At the end of the second iteration, we remove $\xrightarrow{p_1} v_2 \xrightarrow{p_2} T_3$ because its support is lower than $l_\text{min} = 2$, and thus its expansion cannot generate an interesting pattern.
\end{example}

\subsection{Optional and domain-dependent filtering}

After the previous steps, we obtain a feature set $\mathcal{F}$ containing interesting neighbors, paths, and path patterns.
These features have been mined without taking into account domain constraints known to experts.
We propose to apply domain-dependent filtering on $\mathcal{F}$ with parameter $m$.
Such filters reduce the size of $\mathcal{F}$ and integrate interestingness constraints based on expert knowledge.

\begin{example}
	To classify drugs causing or not a side effect, experts may want to focus on features containing a biological pathway, a gene or a GO class, or a MeSH class.
	Therefore, we propose three atomic filters, only keeping neighbors, paths, and path patterns containing at least a pathway ($m=\mtt{p}$), a gene or a GO class ($m=\mtt{g}$), or a MeSH class ($m=\mtt{m}$).
	Such atomic filters can be combined to form disjunctive filters.
	For example, the $m=\mtt{pg}$ filter keeps features from $\mathcal{F}$ containing at least a pathway \emph{or} a gene or a GO class.
	When a filter is applied to a neighbor, this neighbor must be, \textit{e.g.}, a pathway for the \texttt{p} filter.
	When a filter is applied to a path or a path pattern, it means that one of its individuals / ontology classes must be, \textit{e.g.}, a pathway for the \texttt{p} filter.
\end{example}

This domain-dependent filtering is similar to approaches that generalize association rules and prune those that involve some specified ontology classes~\cite{dominguesR11,marinicaG10}.

\section{Experimental setup}
\label{section:experiments}

To illustrate our approach, we will address the following task: from a knowledge graph, mine a set of features to classify drugs depending on whether they cause a specific side effect.

We explore PGxLOD\footnote{\url{https://pgxlod.loria.fr}}~\cite{monninLHRTJNC19}, a knowledge graph that aggregates several sets of Linked Open Data (LOD) describing drugs, phenotypes, and genetic factors: PharmGKB, ClinVar, DrugBank, SIDER, DisGeNET, and CTD.
This aggregation may lead to features combining units from several LOD sets.
Indeed, LOD sets may contain different and incomplete knowledge. 
Their combined use then enables leveraging a greater amount of knowledge, where some LOD sets complete information provided by others.
This asks for a canonical knowledge graph as described in Subsection~\ref{subsection:canonicalization}.
For instance,  it is possible to complete the knowledge related to a drug described in PharmGKB if it is linked with an \texttt{owl:sameAs} arc to the same drug described in DrugBank.
This constitutes the key interest in combining LOD sets in knowledge discovery and data mining tasks, as discussed by Ristoski and Paulheim~\cite{ristoskiP16}.

We will use the following data sets that comprise positive ($\oplus$) and negative ($\ominus$) drug examples:
\begin{dataset}[Drug Induced Liver Injury (DILI)~\cite{chen16}] It is formed by 1,036 drugs in 4 classes: ``most DILI concern'' (192 drugs), ``ambiguous DILI concern'' (254 drugs), ``less DILI concern'' (278 drugs), and ``no DILI concern'' (312 drugs).
	We mapped these drugs from their PubChem identifiers to identifiers from PharmGKB, otherwise DrugBank, otherwise KEGG, resulting in the set of seed vertices $\dili = \pdili \cup \ndili$ such that:
	\begin{itemize}
		\item $|\pdili| = 146$ drugs (118 from PharmGKB, 17 from DrugBank, and 11 from KEGG). The positive drug examples are from the ``most DILI concern'' class.
		\item $|\ndili|= 224$ drugs (206 from PharmGKB, 9 from DrugBank, and 9 from KEGG). The negative drug examples are from the ``no DILI concern'' class. 
	\end{itemize}
\end{dataset}
\begin{dataset}[Severe Cutaneous Adverse Reactions (SCAR)\footnotemark] \footnotetext{\url{http://www.regiscar.org/}} It is formed by 874 drugs in 5 classes: ``very probable'' (18 drugs), ``probable'' (19 drugs), ``possible'' (94 drugs), ``unlikely'' (697 drugs), and ``very unlikely'' (46 drugs).
	We mapped these drugs from their PubChem identifiers to identifiers from PharmGKB, otherwise DrugBank, otherwise KEGG, resulting in the set of seed vertices $\scar = \pscar \cup \nscar$ such that:
	\begin{itemize}
		\item $|\pscar|= 102$ drugs (100 from PharmGKB and 2 from DrugBank). The positive drug examples are from the ``very probable'', ``probable'', and ``possible'' classes.
		\item $|\nscar| = 290$ drugs (286 from PharmGKB and 4 from DrugBank). The negative drug examples are from the ``unlikely'' and ``very unlikely'' classes.
	\end{itemize}
\end{dataset}

We implemented our approach in Python\footnote{\url{https://github.com/pmonnin/kgpm}}.
We used a server with 700~GB of RAM and the following parameter values $k \in \{1,2,3,4\}$, $t \in \{1,2,3\}$, $d = 500$, $u = \mtt{false}$, $l_{\text{min}} = 5$, $l_{\text{max}} = +\infty$, and $m \in \{\mtt{p}, \mtt{g}, \mtt{m}, \mtt{pg}, \mtt{pgm}\}$.
It should be noted that $k = 4$ was only tested with $t = 1$ because of memory issues caused by the high number of generated features.

Statistics about the features are detailed for $k=3$, $t=3$ and $k=4$, $t=1$ in Table~\ref{table:feature-results} and discussed in the next section.
We obtained the features associated with the DILI data set under $k = 3$, $t = 3$ in approximately 1 hour. However, computing the features with $k = 4$, $t = 1$ on the same data set required 4 days and 380~GB of RAM.

\begin{table}
	\centering
	\caption{Number of features mined for the two considered data sets for parameters $d = 500$, $u =\mtt{false}$ before and after applying $l_{\text{min}} = 5$, $l_{\text{max}} = +\infty$, and $m=\mtt{pgm}$.
		Path features are always computed considering at least $l_{\text{min}}$ to avoid combinatorial explosion.
		We display the number of path features generated (gen.) during the exploration and the number of path features ultimately in $\mathcal{F}$ after keeping the most specific and applying limit constraints.
		Types are not considered as features but are displayed to illustrate the possible combinatorial explosion in path patterns.
		Statistics about the full neighborhood are given as comparison.}
	\begin{tabular}{clcccc}
		\toprule
		& & \multicolumn{2}{c}{DILI} & \multicolumn{2}{c}{SCAR} \\
		& & $k=3$, $t=3$ & $k=4$, $t=1$ & $k=3$, $t=3$ & $k=4$, $t=1$ \\
		\midrule
		\multirow{2}*{Before $l_{\text{min}}$ and $l_{\text{max}}$} & Neighbors & 175,652 & 628,681 & 179,694 & 639,050 \\
		& Types & 13,580 & 18,940 & 13,677 & 18,526 \\
		\midrule
		\multirow{4}*{After $l_{\text{min}}$ and $l_{\text{max}}$} & Neighbors & 71,560 & 281,657 & 90,690 & 312,029 \\
		& Types & 7,372 & 12,421 & 8,800 & 13,177\\
		& Path features in $\mathcal{F}$ & 790,605 & 10,169,975 & 1,146,585 & 10,729,002 \\
		& Path features gen. & 20,145,635 & 251,791,519 & 29,011,996 & 255,672,772\\
		\midrule
		\multirow{2}*{After $m$} & Neighbors & 4,069 & 31,804 & 4,214 & 33,361 \\
		& Path features in $\mathcal{F}$ & 102,674 & 2,291,846 & 147,936 & 2,400,642\\
		\midrule
		\multirow{3}*{\makecell{Full neighborhood\\($d = 500$)}} & Neighbors & \multicolumn{2}{c}{2,419,957} & \multicolumn{2}{c}{2,419,920} \\
		& Types & \multicolumn{2}{c}{51,477} & \multicolumn{2}{c}{51,472} \\
		& Reached at $k$, $t$ & \multicolumn{2}{c}{$k = 23$, $t = 21$} & \multicolumn{2}{c}{$k = 23$, $t = 21$} \\
		\midrule
		\multirow{3}*{\makecell{Full neighborhood\\($d = +\infty$)}} & Neighbors & \multicolumn{2}{c}{5,488,531} & \multicolumn{2}{c}{5,488,510} \\
		& Types & \multicolumn{2}{c}{53,486} & \multicolumn{2}{c}{53,484} \\
		& Reached at $k$, $t$ & \multicolumn{2}{c}{$k = 19$, $t = 21$} & \multicolumn{2}{c}{$k = 20$, $t = 21$}\\
		\bottomrule
	\end{tabular}
	\label{table:feature-results}
\end{table}

\section{Results and discussion}
\label{section:discussion}

The first two lines of Table~\ref{table:feature-results} show the number of neighbors and types reachable  before applying support limits.
Enforcing these limits constitutes a first reduction of these numbers, thus reducing the memory and computational footprints.
Here, numbers are approximately divided by 2.
The mining of path features always relies on $l_{\text{min}}$, and thus it is not possible to count the number of all possible paths and path patterns.
However, we show the number of path features generated during the mining, which already illustrates the combinatorial explosion.
Enforcing support constraints and removing redundant generalizations allow to reduce their number in $\mathcal{F}$ (here, approximately by 20).
Finally, the domain-dependent filtering defined by $m$ also radically scales down the number of features ultimately output. 
However, this filtering only happens as post-processing and does not alleviate the scalability issues arising during the mining of patterns.

We observe a drastic increase in the number of neighbors and path features alongside $k$, which highlights the scalability issues of mining large knowledge graphs.
Considering additional levels in ontology hierarchies by increasing $t$ also multiplies the number of path features.
To illustrate, with 700~GB of RAM, we could not set $t$ to values greater than $1$ for $k = 4$.
Consequently, there is still room for improvements in terms of memory consumption, \textit{e.g.}, through an efficient storing of paths and path patterns.
These future improvements could enable to consider the full neighborhood of seed vertices, which is reached for greater values of $k$ and $t$ and involves a far greater amount of vertices.
For example, for the DILI data set, the full neighborhood is reached for $k = 23$ and $t = 21$ (for $d = 500$), or $k = 19$ and $t = 21$ (when the degree constraint is disabled).
The full amount of reachable neighbors is 4 times ($d = 500$) or 9 times greater ($d = +\infty$) than with $k = 4$.

The values of parameters depend on the objectives and domain knowledge of the analyst guiding the mining process, especially for blacklists and support thresholds.
However, metrics about the knowledge graph may provide guidance.
Indeed, statistics about node degrees can help to find a trade-off between exploration and combinatorial explosion with parameter $d$.
Similarly, the depth of class hierarchies influences the value of $t$.
For example, general classes may not be of interest to the analyst, thus reducing $t$.
The parameter $k$ can be set by considering the graph diameter.
As it is common in mining processes, iterations may be required to find the best configuration.
Regarding $m$, it is for now hard-coded and only suitable to some biomedical applications.
Inspired by ontologies that allow to interactively define mining workflows~\cite{ristoskiP16jws}, we could adapt this parameter to other applications by proposing such an interactive definition.

When manually reviewing the output features, we noticed multiple path features across the aggregated LOD sets.
This is made possible by aggregating and canonicalizing multiple LOD sets in the knowledge graph.
This result particularly illustrates one of the fundamental aspects of Linked Open Data: the combination of different data sets enables to go beyond their original purposes and coverage.
However, it is clear that combining LOD sets leads to bigger knowledge graphs, exacerbating the scalability issues.

Regarding our approach, we only canonicalize vertices, \textit{i.e.}, individuals and ontology classes.
Nevertheless, predicates used on arcs can also be identified as identical, leading to a canonicalization of arcs.
In this context, we could benefit from matching approaches, such as PARIS~\cite{suchanekAS11}. 
By identifying identical classes, predicates, and individuals, these matching approaches could further improve the canonicalization and, therefore, increase the number of common features between seed vertices from different data sets.
Similarly, we could consider literals and arcs incident to literals that were purposely discarded here.
However, the canonicalization of literals raises several challenging issues due to their heterogeneity in terms of syntactic variations, unit measures, and the precision of numerical values.
Other reasoning mechanisms and semantics associated with Semantic Web standards could be taken into account.
For example, predicates can be defined as transitive, and thus the canonical knowledge graph could also result from their transitive closure.

Regarding the modeling of path patterns, we could generalize paths with both the hierarchy of classes and the hierarchy of predicates.
In addition to keeping the most specific patterns, we could use other metrics to further reduce their redundancy (\textit{e.g.}, approaches relying on hierarchies~\cite{ristoskiP14,dAmatoSF08} and extents of ontological classes~\cite{dAmatoSF08}).
This could also reduce the number of generated patterns, therefore improving the scalability of the mining approach.
Neighbors could be enriched with the distance between them and seed vertices, which would correspond to the generalized paths of KGPTree~\cite{vandewiele19}.
We could also use other approaches than binary features (\textit{e.g.}, counting~\cite{vriesR13,vriesR15}, relative counting~\cite{ristoskiP14pkdd}).
More importantly, it remains to test our mined features within a complete classification task to measure the influence of $k$, $t$, and the three kinds of features (neighbors, paths, and path patterns).

\section{Conclusion}
\label{section:conclusion}

In this preliminary study, we addressed the scalability issues associated with the task of mining neighbors, paths, and path patterns from a knowledge graph and a set of seed vertices. 
We proposed a method for tackling these issues, which we illustrated by mining a real-world knowledge graph.
Our results highlight the importance of considering the scalability of approaches when mining features from ever-growing knowledge graphs.
Our work alleviates part of the computational cost (time and memory) of mining paths and path patterns but also reveals the need for a further reduction.
Such future research works could enable the modeling of more complex path patterns, for example, considering the hierarchy of predicates.

\bibliographystyle{unsrt}  
\bibliography{references}  

\end{document}